\documentclass[12pt,preprint]{aastex}

\usepackage{subfigure}

\slugcomment{}

\shorttitle{} \shortauthors{Ling Zhang Tang}

\begin{document}

\title{Determining the Distance of Cyg X-3 with its X-ray Dust Scattering Halo}

\author{Zhixing Ling\altaffilmark{1}, Shuang Nan
  Zhang\altaffilmark{1,2,3}, and Shichao Tang\altaffilmark{1}}

\altaffiltext{1}{Department of Physics and Tsinghua Center for
  Astrophysics, Tsinghua University, Beijing 100084, China;
  lingzhixing@tsinghua.org.cn, zhangsn@tsinghua.edu.cn,
  tsc02@mails.tsinghua.edu.cn} \altaffiltext{2}{Key Laboratory of
  Particle Astrophysics, Institute of High Energy Physics, Chinese
  Academy of Sciences, P.O. Box 918-3, Beijing 100049, China}
\altaffiltext{3}{Physics Department, University of Alabama in
  Huntsville, Huntsville, AL 35899, USA}

\begin{abstract}

Using a cross-correlation method, we study the X-ray halo of Cyg
X-3. Two components of dust distributions are needed to explain the
time lags derived by the cross-correlation method. Assuming the
distance as 1.7 kpc for Cyg OB2 association (a richest OB
association in the local Galaxy) and another uniform dust
distribution, we get a distance of $7.2^{+0.3}_{-0.5}$ kpc (68$\%$
confidence level) for Cyg X-3. When using the distance estimation of
Cyg OB2 as 1.38 or 1.82 kpc, the inferred distance for Cyg X-3 is
$3.4^{+0.2}_{-0.2}$ or $9.3^{+0.6}_{-0.4}$ kpc respectively. The
distance estimation uncertainty of Cyg X-3 is mainly related to the
distance of the Cyg OB2, which may be improved in the future with
high-precision astrometric measurements. The advantage of this
method is that the result depends weakly on the photon energy, dust
grain radius, scattering cross-section, and so on.

\end{abstract}


\section{Introduction}
The X-ray dust scattering halo was first discussed by Overbeck in
1965. From then on, many authors developed the theory of scattering.
However, there was no direct evidence for this phenomenon until
1980s when Rolf (1983) first observed the X-ray halo by analyzing
the data of GX339-4 with the imaging proportional counter (IPC)
instrument onboard the \textit{Einstein X-ray Observatory}. There
are two groups of methods to study the X-ray scattering halo. The
first is evaluating the halo surface brightness distribution around
the point source. During the past quarter century, X-ray halos can
be found in the data of \textit{Einstein}, \textit{ROSAT,
XMM-Newton, Chandra } and \textit{Swift}. Predehl \& Schmitt (1995)
analyzed the data of \textit{ROSAT} and found a strong correlation
between the visual extinction and the hydrogen column density of 25
point sources. Vaughan et al. (2004, 2006) found ring structures in
two gamma-ray burst (GRB) observations with \textit{XMM-Newton} and
\textit{Swift}. Smith, Edgar \& Shafer (2002) reported the halo of
GX 13+1 between 50$^{\prime \prime}$ and 600$^{\prime \prime}$ with
the data of Advanced CCD Imaging Spectrometer (ACIS) onboard the
\textit{Chandra X-ray Observatory}. Yao et al. (2003) determined
that the halo of Cyg X-1 is as close to the point source as
1$^{\prime \prime}$, using a reconstruction method with the data of
Continuous Clocking Mode of ACIS. Xiang, Zhang \& Yao (2005)
reconstructed the halo's surface brightness of 17 bright sources and
deduced the dust distribution along the line of sight (LOS) with the
data from ACIS-S array.

The other way is to study the effect of delay and broadening of the
light curve. Tr$\mathrm{\ddot{u}}$mper \&
Sch$\mathrm{\ddot{o}}$nfelder (1973) first proposed to use the delay
and smearing property to determine the distances of the X-ray
sources. Predehl et al. (2000) first used the delay property in
determining the distance of Cyg X-3 with the data of ACIS. Hu, Zhang
\& Li (2004) developed a method of using the power density spectra
to determine the distances of X-ray sources. Xiang, Lee \& Nowak
(2007) used the delay property to determine the distance of 4U
1624-490. Vaughan et al. (2004, 2006) evaluated the distances of
some dust molecular clouds by the delayed ring structures in two GRB
observations with \textit{XMM-Newton} and \textit{Swift}. Thompson
\& Rothschild (2008) used the eclipse data determining the distance
of Cen X-3 to be 5.7 $\pm$ 1.5 kpc. Ling et al. (2009) first used
the cross-correlation method to study the light curves of the X-ray
halo of Cyg X-1. They found obvious time lag peaks in the
cross-correlation curves. All those peaks revealed a dust
concentration at a distance of 1.76 kpc from us.

Actually, the first method uses the halo surface brightness
distribution to study the dust distribution and dust model, whereas
the second method studies the dust distribution and source distance.
The goals from each type of study are usually different. Both
techniques require careful point-spread function (PSF) subtraction
during the analysis.

In this work, we re-analyze the data of Cyg X-3 with the
cross-correlation method described in Section 2. We discuss the
multiple scatterings in Section 4. After obtaining the
cross-correlation curves from 15$^{\prime \prime}$ to 80$^{\prime
\prime}$, we use two dust distributions to explain the time lags.
Assuming the distance as 1.7 kpc for Cyg OB2 and another uniform
dust distribution, we get a distance of $7.2^{+0.3}_{-0.5}$ kpc
(68$\%$ confidence level) for Cyg X-3 in Section 5. When using the
distance estimation of Cyg OB2 as 1.38 or 1.82 kpc, the inferred
distance for Cyg X-3 is $3.4^{+0.2}_{-0.2}$ or $9.3^{+0.6}_{-0.4}$
kpc respectively. We summarize our results in Section 6.

\section{Method and Data preparation}

The details of X-ray dust scattering can be found in Van de Hulst
(1957), Overbeck (1965), Tr$\mathrm{\ddot{u}}$mper \&
Sch$\mathrm{\ddot{o}}$nfelder (1973), and Smith \& Dwek (1998).
Here, we just show some equations used in this work.

As shown in Figure 1, an X-ray source is located at a distance of
$\mathit{D}$. The dimensionless number $\mathit{x}$ is the ratio of
the distance of scattering and that of the source from us. So the
lag time of scattered photons at $\mathit{x}$ can be expressed as

\begin{equation}
t_\mathrm{Delay}(\phi,x)=\left(\frac{x}{\cos\phi}+\sqrt{(1-x)^2+(x\tan\phi)^2}-1\right)\times
\frac{D}{c}.
\end{equation}

Let $\mathit{I(t)}$ denote the observed flux of the source at $x =
0$, the observed halo intensity at different observational angle
$\phi$ is given by

\begin{equation}
H(\phi,t)=\int_{0}^{1}{{\mathrm{d}x}}\times{D}\times
\frac{I(t-t_\mathrm{Delay}(\phi,x))\times{\rho(x)}}{(1-x)^2}
\times\frac{d\sigma(\theta)}{d\Omega},
\end{equation}
where the scattering cross section $\frac{d\sigma(\theta)}{d\Omega}$
depends on the energy of the X-ray photon and the radius of the dust
grain. $\rho(x)$ is the density (in units of $\mathrm{cm^{-3}}$) of
the dust grain at $\mathit{x}$. If $\mathit{I(t)}$ equals to a delta
function, equation 2 would evolve to be a response function of a
delta function (we denote this function by $\mathit{R(\phi,t)}$
hereafter). This is the situation of a GRB, hence equation 2 would
be an ideal observed light curve of a halo at a given observational
angle $\mathit{\phi}$. For the other situations, the light curve of
the halo at angle $\phi$ equals to the convolution of
$\mathit{I(t)}$ and $\mathit{R(\phi,t)}$. This process can also be
understood by Figure 1. The light curve of the halo at observational
angle $\phi$ is an integral effect of scattering from the dust near
the observer to the source. From this process, the light curve of
the halo would be lagged and smeared from the light curve of the
source.

We can study the delay property directly with the cross-correlation
method (Ling et al. 2009). The definition of cross-correlation
coefficient is given by
\begin{equation}
c(\Delta t)=\frac{1}{N-\arrowvert {\Delta
t}\arrowvert}\sum_{t=0}^{N-\arrowvert {\Delta t} \arrowvert-1}  (
L_{h}(t+\Delta t) -\mu_h) (L_{s}(t)-\mu_s), \label{form:ccf}
\end{equation}
here $\mathit{L_{s}}$ and $\mathit{L_{h}}$ are the light curves of
the X-ray source and halo (at a given observational angle $\phi$) in
the same energy band. $\mu_s$ and $\mu_h$ are the average values of
$\mathit{L_{s}}$ and $\mathit{L_{h}}$ respectively.

\begin{table}[!h]
\tabcolsep 5mm \caption{Observations of Cyg X-3 we used.}
\begin{center}
\begin{tabular}{|c|c|c|c|}
\hline ObsID &Exposure (ks)&Start date&Instrument
\\
\hline
1456 &20.8   &1999-10-20 &ACIS-S \\
425  &21.53  &2000-04-04 &ACIS-S\\
426  &18.32  &2000-04-06 &ACIS-S\\
\hline
\end{tabular}
\end{center}
\end{table}

\section{Data extraction and Result}

Three observational data sets, listed in table 1, are used in our
analysis. The CIAO version 3.4 and CALDB version 3.4.1 are used to
process the observational data. The main process in this study is
similar to the way described in Ling et al. (2009). Here, we just
give a brief description. First, we divide the observational data
into three energy bands: below 3 keV (band I), 3 keV-5 keV (band
II), and above 5 keV (band III).  Second, we get the light curve of
the point source by the photons in the region of the streak. Third,
the light curve of the halo at each bin (the bin width is 5$^{\prime
\prime}$) of observational angle is obtained from the photons of
different annuli around the point source. All background
contributions have been excluded for all the light curves above.
After extracting all light curves, we make cross-correlation curves
between the light curves of the halo and the light curve of the
source in Figure 2-4. The top curve in each panel is the
autocorrelation of the light curve of the source; all other curves
are the cross-correlation curves from 15$^{\prime \prime}$ to about
90$^{\prime \prime}$ with a step of 5$^{\prime \prime}$. For
clarity, the cross-correlation coefficients have been lowered by a
same amount successively for each curve. The auto-correlation curves
have a peak because of the intrinsic variations of 4.8 hr of the
source. The peaks of the cross-correlation curves of the halo below
50$^{\prime \prime}$ lagged a little from the center; however the
peaks of 60$^{\prime \prime}$ and 65$^{\prime \prime}$ advanced
ahead of the center obviously. The relationship between the lag time
and the observational angle of Cyg X-3 is quite different from that
of Cyg X-1 (Ling et al. 2009), of which the lag times moved longer
gradually with the increasing angles. Therefore, the lag times of
Cyg X-3 cannot be explained by the scattering of a single dust wall,
as suggested by the data of Cyg X-1. The moving effect is not clear
in band III because of the low cross section of scattering for high
energy photons. Because of its low count rate, the curves of
ObsID1456 show peaks less clearly than the other two observations.

In Figure 2-4, the cross-correlation curves are contaminated by the
instrument because of the PSF effect. As described in Ling et al.
(2009), ChaRT and Marx are used to simulate the contaminated factor
of the PSF. After obtaining those factors of contamination, we could
get a cleaned cross-correlation curve. The cleaned cross-correlation
curves are shown in Figure 5-8. By fitting the peaks with a simple
Gaussian function, we get the time lag at each angle. The dashed
lines in each panel show the fitting result of a Gaussian function.
Those four figures show the result of all of the data that have
obvious lag in the cross-correlation curves. We list the time lags
obtained from the cross-correlation curves in Table 2. The curves of
ObsID1456 have no peaks at angles greater than 50$^{\prime \prime}$
because of its lower count rate. The band I of this observation also
has no peak (as can be seen in Figure 4 directly). From those
figures, we find that the lag time of the cross-correlation method
of band I are similar with the uncleaned cross-correlation curves of
Figure 2 and 3. This also happened in the study of Cyg X-1 (Ling et
al. 2009). The reason is that the PSF of the low energy band is
narrower than the high energy band.

In the panel of 65$^{\prime \prime}$ in band II of ObsID425 (Figure
6, left side), there are two obvious peaks in the cleaned
cross-correlation curves. The left peak with a time lag of about 10
ks and the right peak with a time lag of about $-$6 ks. The interval
between those two peak is about 17 ks, which equals to the period of
the light curve of source, i.e., 4.8 hr. This result confirms that
the time lag derived by cross-correlation method is real but not
noise. There are many other curves that show this property: from
50$^{\prime \prime}$ to 65$^{\prime \prime}$ in band II of ObsID425
and ObsID426 show two peaks. The curve of band III is not clear
because the low efficiency of the scattering. Consequently, this
phenomenon confirms that the time lags of 50$^{\prime
\prime}$-65$^{\prime \prime}$ we derived from the cross-correlation
curves are reliable. For the data of 80$^{\prime \prime}$ and
85$^{\prime \prime}$, there is only one lag time from band I of
ObsID425; thus we do not use those data for further analysis in the
following section.

\begin{table}[!h]
\tabcolsep 0.1mm \caption{Time lag of cross-correlation curve (in
Units of Second).}
\begin{center}
\begin{tabular}{|c|c|c|c|c|c|c|c|c|c|}
\hline Angle&ID425&ID425&ID425&ID426&ID426&ID426&ID1456&ID1456&mean\\
\hline
Band   &BandI&BandII&BandIII&BandI&BandII&BandIII&BandII&BandIII&-\\
\hline
15$^{\prime \prime}$&3467&1424&975&5303&2281&1513&410&889&2032$\pm$621\\
20$^{\prime \prime}$&2011&931&915&146&677&882&82&390&754$\pm$233 \\
25$^{\prime \prime}$&3508&2241&1548&4081&2456&2391&1109&426&2220$\pm$459\\
30$^{\prime \prime}$&2121&1918&1483&3906&2873&2102&160&-&2080$\pm$477\\
35$^{\prime \prime}$&5069&3028&2327&4899&3792&2956&8043&546&3833$\pm$852\\
40$^{\prime \prime}$&3723&3840&3070&4937&4441&3617&1935&-&3652$\pm$643\\
45$^{\prime \prime}$&4844&3402&1903&2152&5789&3116&2613&-&3403$\pm$733\\
50$^{\prime \prime}$&5604&2839&1862&4058&3076&1657&-&-&2735$\pm$667\\
55$^{\prime \prime}$&5866&4670&1885&4727&5046&3965&-&-&3745$\pm$616\\
60$^{\prime \prime}$&6834&6797&4240&4930&5893&4669&-&-&4775$\pm$505\\
65$^{\prime \prime}$&9558&10450&-&9129&9427&11617&-&-&10036$\pm$516\\
70$^{\prime \prime}$&14617&16297&-&-&13829&15084&-&-&14957$\pm$620\\
75$^{\prime \prime}$&19370&15961&-&-&-&18833&-&-&18055$\pm$1397\\
80$^{\prime \prime}$&24833&-&-&-&-&-&-&-&-\\
85$^{\prime \prime}$&32488&-&-&-&-&-&-&-&-\\
\hline
\end{tabular}
\end{center}
\end{table}

\section{Multiple Scatterings}
Before the analysis of the data of table 2, we estimate the
influence of the multiple scatterings. Because of its high hydrogen
column density, the multiple-scattered photons may contaminate the
observed light curve of the halo.  The details of multiple
scatterings can be found in Mathis $\&$ Lee (1991). The conclusion
in their study was that the single scattering dominates at small
angles ($\theta < 60^{\prime \prime}$ for $\tau_\mathrm{sca} = 2$ at
\emph{E} = 1 keV).

A Monte Carlo simulation is conducted to estimate the effect of
multiple scatterings. The parameters we used are $\tau_\mathrm{sca}
= 2$ at \emph{E} = 1 keV and \emph{a} = 0.1 $\mu$m. The simulated
fractions of multiple-scattered photons to the total halo photons
are shown in Figure \ref{fig:multi}. The solid line shows the
fraction of multiple-scattered photons for band I (below 3 keV). The
dotted line shows the fraction of multiple-scattered photons for
band II (3-5 keV). The dashed line shows the fraction of
multiple-scattered photons for band III (above 5 keV). From Figure
\ref{fig:multi}, in band I the fraction of multiple-scattered
photons is 8$\%$, 13$\%$, and 15$\%$ at observational angles of
30$^{\prime \prime}$, 60$^{\prime \prime}$, and 80$^{\prime
\prime}$, respectively. Compared with the uncertainties for the lag
times derived in table 2, the influence of multiple-scattered
photons can be ignored in our analysis.

\section{Analysis and Discussion}
\subsection{Uniform dust distribution}
Predehl et al. (2000) analyzed the data of Cyg X-3 previously. They
compared the light curve of the halo within about 10$^{\prime
\prime}$ and the light curve of the point source. They found a lag
of about 2 ks in the light curve directly. Assuming a uniform dust
distribution, they got a distance of $9^{+4}_{-2}$ kpc for Cyg X-3.
In our study, three observations are used to determine the lag time
up to 75$^{\prime \prime}$. Figure \ref{fig:fit} shows the result of
lag time at each observational angle. From our result, the time lags
of 60$^{\prime \prime}$-75$^{\prime \prime}$ show significant
increase compared to the other time lags.

First, we try to fit the data with the same model described by
Predehl et al. (2000). A uniform dust distribution is used to
produce an $\mathit{R(\phi, t)}$. By convoluting the
$\mathit{R(\phi, t)}$ with an ideal sinusoidal wave, we get a
simulated light curve of halo at observational angle $\phi$. The
dust model we used here is proposed by Mathis, Rumpl \& Nordsiech
1977 (MRN hereafter). We show the process and the result in Figure
\ref{fig:convection}. The top panel shows a sinusoidal wave,
representing the light curve of the source. The middle panel shows
three curves of $\mathit{R(\phi, t)}$ in logarithm scale. The solid
line shows the $\mathit{R(\phi, t)}$ with the parameter of
$\mathit{D} = 2$ kpc and $\phi = $15$^{\prime \prime}$, the dotted
line shows the $\mathit{R(\phi, t)}$ of $D = 10$ kpc and $\phi$ =
15$^{\prime \prime}$, and the dashed line shows the $\mathit{R(\phi,
t)}$ of $D = 2$ kpc and $\phi = 50$$^{\prime \prime}$. The bottom
panel shows the result of convolution of the sinusoidal wave with
these three different $\mathit{R(\phi, t)}$. These three curves are
treated as the light curves of halo at different angles. The
vertical axis of the bottom panel is normalized to unity for
clarity. The three simulated light curves of the halo show different
magnitude but have the same lag time. This result shows that if the
distance of the source exceeds 2 kpc, the time lag between the light
curve of the halo and that of the source will be a constant: around
3.5 ks (less than $T$/4 of the sinusoidal wave). This is also the
situation for any angle larger than 15$^{\prime \prime}$. The main
reason is that the profile of $\mathit{R(\phi, t)}$ is much longer
than the period of the sinusoidal wave.

\subsection{\textbf{Single dust wall model: Cyg Ob2 association}}
From the discussion of Cyg X-1 (Ling et al. 2009), we found that the
dust distribution is quite nonuniform toward Cyg X-1: a dust
concentration at a distance of 2.0 kpc $\times$ (0.876 $\pm$ 0.002)
from the Earth is found. Thus, we try to find some dust cloud around
the region of Cyg X-3 to explain the time lag of 65$^{\prime
\prime}$ and the greater than 65$^{\prime \prime}$.

A likely candidate for the dust concentration is the Cyg OB2
association. Cyg OB2 is one of the richest OB associations in the
local Galaxy; it houses many of the hottest and most luminous stars
known in our Galaxy. Cyg X-3 lies in the field of Cyg OB2
(Kn$\mathrm{\ddot{o}}$dlseder 2003). Hutchings (1981) assumed an
absolute distance modulus (m$-$M) = 10.7, converting to $\mathit{d}$
= 1.38 kpc. Humphreys (1978) adopted $d$ = 1.82 kpc, while
Torres-Dodgen et al. (1991) and Massey \& Thompson (1991) determined
$d$ = 1.74 kpc ((m - M) = 11.2). Kn$\mathrm{\ddot{o}}$dlseder (2000)
assumed a distance $d =$ 1.7 kpc. In this work, a distance of 1.7
kpc for Cyg OB2 is used in the following analysis.

Assuming the distance of Cyg X-3 to be 10 kpc, the time lag of
65$^{\prime \prime}$ reveals that a dust concentration exists at a
distance of about 2 kpc. This result is consistent with the distance
of Cyg OB2. Then taking the distance of Cyg OB2 to be 1.7 kpc, we
use a single dust wall model to fit the observed time lag. The
result is shown in Figure \ref{fig:fit}, in which the dashed line
shows the shows a model assuming a dust wall 1.7 kpc from the Sun
and a distance of 5 kpc to the source. The dotted line shows the
same model with a distance of 10 kpc to the source. Those two curves
cannot fit the observed lags.

\subsection{Uniform distribution plus dust wall}
Combining the results of Section 5.1 and 5.2, we propose to use two
components to fit the observed time lags. We divide the observed
time lags into two parts: below 60$^{\prime \prime}$ and above
65$^{\prime \prime}$. The first part mainly comes from the component
of a uniform dust distribution and the second part is mostly due to
a dust concentration of Cyg OB2 at the distance of around 1.7 kpc,
reflecting that the small angle halo tends to explore the dust near
the source (Mathis $\&$ Lee 1991). The new two-components dust
distribution model needs two parameters to fit the observed time
lags. The first parameter is the distance $D$ of the point source
and the other parameter is the fraction of the dust concentrated in
Cyg OB2. The solid line of Figure \ref{fig:fit} shows the
best-fitting result. From this result, we get a distance of
$7.2^{+0.3}_{-0.5}$ kpc (68$\%$ confidence level) of Cyg X-3. The
fraction of dust concentrated in Cyg OB2 is $7^{+1.0}_{-0.5}\%$
(68$\%$ confidence level). The uncertainty of the distance is
calculated by $\Delta \chi^2= 2.3$ (Avni 1976). At the same time, we
refit the data with the distance assumption of Cyg Ob2 to be 1.38
and 1.82 kpc away from us. The fitting results are
$3.4^{+0.2}_{-0.2}$ and $9.3^{+0.6}_{-0.4}$ kpc, respectively.

Alternatively, a new way is proposed to give a range for the
distance of Cyg X-3. We define the response function of a uniform
dust distribution with $\mathit{R_{u}(t)}$, and the response
function of the dust wall with $\mathit{R_{w}(t)}$ for simplicity.
Then, the light curve of the halo is given by

\begin{equation}
H(t) = L(t) \otimes (R_{u}(t)+R_{w}(t)), \label{form:conv1}
\end{equation}

here $\otimes$ stands for convolution. Equation 4 can be decomposed
to
\begin{equation}
H(t) = L(t) \otimes R_{u}(t) + L(t) \otimes R_{w}(t).
\label{form:conv2}
\end{equation}

Equation 5 can be understood as the light curve of the halo being
the sum of the two components from the two dust distributions. We
use $H_{u}(t)$ and $H_{w}(t)$ hereafter to identify those two
components.  As pointed out in Section 5.1, the $H_{u}(t)$ may cause
a lag of about 3.5 ks (of course it must also have a period of $T$,
the same as the $L(t)$). The time lag of the $H_{w}(t)$ depends on
the distance of Cyg X-3, but obviously $H_{w}(t)$ has a period of
$T$ too. As a result, the phase of $H(t)$ is related to these two
components: $H_{u}(t)$ and $H_{w}(t)$. As shown in Figure
\ref{fig:sum}, the top panel represents the light curve of $L(t)
\otimes R_{u}(t)$, the middle panel represents the light curve of
$L(t) \otimes R_{w}(t)$. The left part of the middle panel has a lag
of less than $T/2$ with respect to the top panel, and the right part
has a lag of larger than $T/2$ with respect to the top panel. The
bottom panel shows the sum of the above two panels. The arrows show
the lag of the summed curves. The sum of those two components can
cause two possible results: if
\begin{equation}
Lag(H_{u}(t)) = 3.5\ \mathrm{ks} < Lag(H_{w}(t)) <  12\ \mathrm{ks}
= Lag(H_{u}(t))+T/2, \label{form:conv3}
\end{equation}
then
\begin{equation}
3.5\ \mathrm{ks} < Lag(H(t)) < 12\ \mathrm{ks}, Lag(H(t)) <
Lag(H_{w}(t)); \label{form:conv4}
\end{equation}

Alternatively if
\begin{equation}
12\ \mathrm{ks} < Lag(H_{w}(t)) < 12\ \mathrm{ks} + T/2,
\label{form:conv4}
\end{equation}
then
\begin{equation}
12\ \mathrm{ks} < Lag(H(t)) < 12\ \mathrm{ks} + T/2, Lag(H(t)) >
Lag(H_{w}(t)). \label{form:conv4}
\end{equation}

The first situation means that when the lag time of $H_{w}(t)$ is
less than the time lag of $H_{u}(t)$ plus $T$/2, i.e., about 12 ks,
the time lag of $H(t)$ will be between 3.5 and 12 ks. The second
case means that if the time lag of $H_{w}(t)$ is between 12 and 20.5
ks, the time lag of the summed curve will be greater than the time
lag of $H_{w}(t)$. In other words, the second situation means that
the observed time lag of the halo could be longer than the time lag
caused by the dust wall.

Let us apply these results to the observed time lags of Figure
\ref{fig:fit}. The time lag at 65$^{\prime \prime}$ is around 10 ks,
less than 12 ks. Therefore, the time lag of $H_{w}(t)$ must be less
than 12 ks, as we illustrated in Figure \ref{fig:distance}. In
Figure \ref{fig:distance}, the solid line is the lag time at
60$^{\prime \prime}$, the dotted line is the lag time at 65$^{\prime
\prime}$, the dashed line is the lag time at 70$^{\prime \prime}$,
and the dashed dotted line is the lag time at 75$^{\prime \prime}$.
With the constraint of 12 ks, the distance of the source must be
greater than 4.5 and 6 kpc by the data of 60$^{\prime \prime}$ and
65$^{\prime \prime}$. The maximum of the distance can be derived by
the data of 70$^{\prime \prime}$, of which the time lag of the dust
wall must exceed 20.5 ks; a distance upper limit of
10 kpc is derived at this angle. The time lag of 75$^{\prime
\prime}$ would give an upper limit of 15 kpc. Using all these
result, we give a range of [6, 10] kpc for Cyg X-3. The best-fit
result of $7.2^{+0.3}_{-0.5}$ kpc is among this range obviously.

\subsection{Halo surface brightness of Cyg X-3}
After obtaining the ratio between the two dust components, we can
predict a halo surface brightness distribution with a dust grain
model. We find that the halo surface brightness distribution
predicted by the MRN dust model and the two-components dust
distribution cannot fit the observed halo surface brightness derived
in Xiang, Zhang \& Yao (2005). In Figure \ref{fig:surface}, the
dotted line shows the predicted surface brightness of a uniform
distribution, and the dashed line shows the predicted surface
brightness caused by Cyg OB2. The total $\mathit{N_{H}}$ used here
is $3.0 \times 10^{22}$ $\mathrm{cm^{-2}}$ (the $\mathit{N_{H}}$
derived by Predehl and Schmitt (1995) is $3.31 \times 10^{22}$
$\mathrm{cm^{-2}}$), and the fraction of the dust in Cygnus OB2 is
7$\%$; clearly the dust wall has almost no influence on the halo
surface brightness. The predicted halo surface brightness of source
cannot fit the observed surface brightness of Cyg X-3 obviously.
Similar discrepancy has been found in our previous work on Cyg X-1
(Ling et al. 2009). To fit the halo surface brightness of angles
smaller than 10$^{\prime \prime}$, we add a new dust component
between \emph{x} = 0.99 and \emph{x} = 1.0. The fitting result is
shown by the solid line in Figure \ref{fig:surface}. The column
density $\mathit{N_{H}}$ of the uniform distribution, the dust wall
and the dust near the source are 7.0 $\pm$ 0.3 $\times 10^{21}$
$\mathrm{cm^{-2}}$, 32.0 $\pm$ 1.7 $\times 10^{21}$
$\mathrm{cm^{-2}}$ and 3.04 $\pm$ 0.06 $\times 10^{21}$
$\mathrm{cm^{-2}}$ respectively. The total $\mathit{N_{H}}$ from the
fitting is consistent with the result of Predehl and Schmitt (1995).
Our fitting shows that the ratio of $N_{H}$ in the dust wall to the
$N_{H}$ of the uniform dust distribution is 4.6, conflicting
significantly from our result of 0.075 derived with the
cross-correlation method which is almost independent of the dust
size distribution model. As a result, we conclude that the MRN dust
model must be modified before it is used in the X-ray regime.


\subsection{Independence of dust grain model}
In Section 5.4, we have shown that the MRN model is not sufficient
in modeling the halo surface brightness distribution along the LOS
of Cyg X-3. However, in the analysis of Section 5.1, the MRN dust
grain model was used to produce $R_{u}(t)$, the response function of
a uniform dust distribution. Here, we address the question whether
our result is dependent of the dust grain model. From equation 2, we
can get $R_{u}(t)$ with different dust radii directly. Then, we
could get a simulated $H_{u}(t)$ for any dust radius. After
comparing the phase of the light curve of source and $H_{u}(t)$, we
get the lag time in different dust radii. Figure \ref{fig:dustmodel}
shows the lag time of $H_{u}(t)$ of 15$^{\prime \prime}$ versus the
distance of the source. The solid line represents the lag time of
$H_{u}(t)$ with a dust radius of 0.005 $\mu$m, the dashed line
represents the lag time of $H_{u}(t)$ with a dust radius of 0.05
$\mu$m, and the dotted line represents the lag time of $H_{u}(t)$
with a dust radius of 0.25 $\mu$m. The lag time approaches to 3.5 ks
when the distance of the source exceeds 5 kpc for any dust radius.
The range of grain radii used in the MRN model is [0.005, 0.25]
$\mu$ m, and the range of grain radii used in the Weingartner $\&$
Drain (2001; WD01) model is also similar. Therefore, the different
$R_{u}(t)$, with different dust grain radii, would produce a lag of
3.5 ks for $H_{u}(t)$. The conclusion is that the result of Section
5.1 is independent with the dust grain model. At the same time, we
fit the lag time and with a single dust grain radius instead of the
MRN dust grain model of Section 5.3. The best-fit result is
$7.2^{+0.2}_{-0.4}$ kpc for 0.25 $\mu$m, $6.1^{+0.4}_{-0.2}$ kpc for
0.05 $\mu$m, and $6.5^{+0.4}_{-0.2}$ kpc for 0.005 $\mu$m. The
uncertainty from the dust model almost equals to the uncertainty of
the statistical error of the distance of Section 4.3. By these
results, we conclude that the distance of Cyg X-3 we derived in this
work  is independent of the dust grain model.

\section{Summary and Discussion}

We applied the cross-correlation method to the light curves of Cyg
X-3 and found the time lag from the cross-correlation curves between
the angles of 15$^{\prime \prime}$ to about 90$^{\prime \prime}$.
The time lags reveal that there are two components of dust
distributions in the LOS toward Cyg X-3: a uniform distribution and
a dust concentration. A likely candidate for the dust concentration
is the Cyg OB2 association. Assuming the distance as 1.7 kpc for Cyg
OB2 \and another uniform dust distribution, we obtain a distance of
$7.2^{+0.3}_{-0.5}$ kpc for Cyg X-3. Multiple scattering makes no
influence for the distance estimation in our analysis. The
systematic uncertainty may come from the uncertainty of the distance
of the Cyg OB2. When using the distance estimation of Cyg OB2 as
1.38 or 1.82 kpc, the inferred distance for Cyg X-3 is
$3.4^{+0.2}_{-0.2}$ or $9.3^{+0.6}_{-0.4}$ kpc, respectively.

As discussed by Predehl et al. 2000, the distance of Cyg X-3 has
been a puzzle for a long time. Dickey (1983) has found a lower limit
of 9.2 kpc using 21 cm wavelength absorption data. Predehl \&
Schmitt (1995) derived 8 kpc as the distance through the galactic
dust layer from their comparison of X-ray scattering and absorption.
A distance of $7.2^{+0.3}_{-0.5}$ kpc is only about 3/4 to the
previous result. For example, the estimation of velocity of the
radio jet of Cyg X-3 decrease from 0.5c to 0.36c (Mart\'{\i} et al.
2001). The new velocity is comparable with SS433, which has a radio
jet velocity of 0.26c (Milgrom 1979). From our discussion, at small
observational angle (below 100) the cross-correlation method is only
weakly dependent of the photon energy, dust grain radius, scattering
cross-section, and so on. Therefore, the time lag derived by this
method rests almost purely on geometry. For Cyg X-3, the distance
estimation uncertainty is mainly related to the distance of the Cyg
OB2 association, which may be improved in the future with
high-precision astrometric measurements.

Consequently, our results can be used to determine the parameter of
the dust grain models in the future, when combined with the spatial
distribution of the X-ray dust scattering halo; currently no dust
grain model can describe simultaneously the time lag and spatial
distribution of X-ray dust scattering halo.

\acknowledgments

We thank Randall K. Smith for providing the model codes. The
anonymous referee is thanked for many constructive comments and
useful suggestions. S.N.Z. acknowledges partial funding support by
the Yangtze Endowment from the Ministry of Education at Tsinghua
University, Directional Research Project of the Chinese Academy of
Sciences under project no KJCX2-YW-T03 and by the National Natural
Science Foundation of China under grant no 10521001, 10733010,
10725313, and by 973 Program of China under grant 2009CB824800.

\begin{figure}
\begin{center}
  \includegraphics[angle=0,scale=.3]{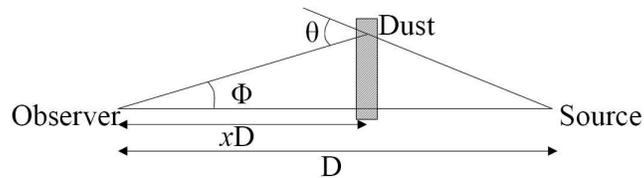}\\
  \caption{X-ray dust scattering geometry.}\label{fig:illu}
\end{center}
\end{figure}

\begin{figure}
\begin{center}
\includegraphics[angle=0,scale=0.7]{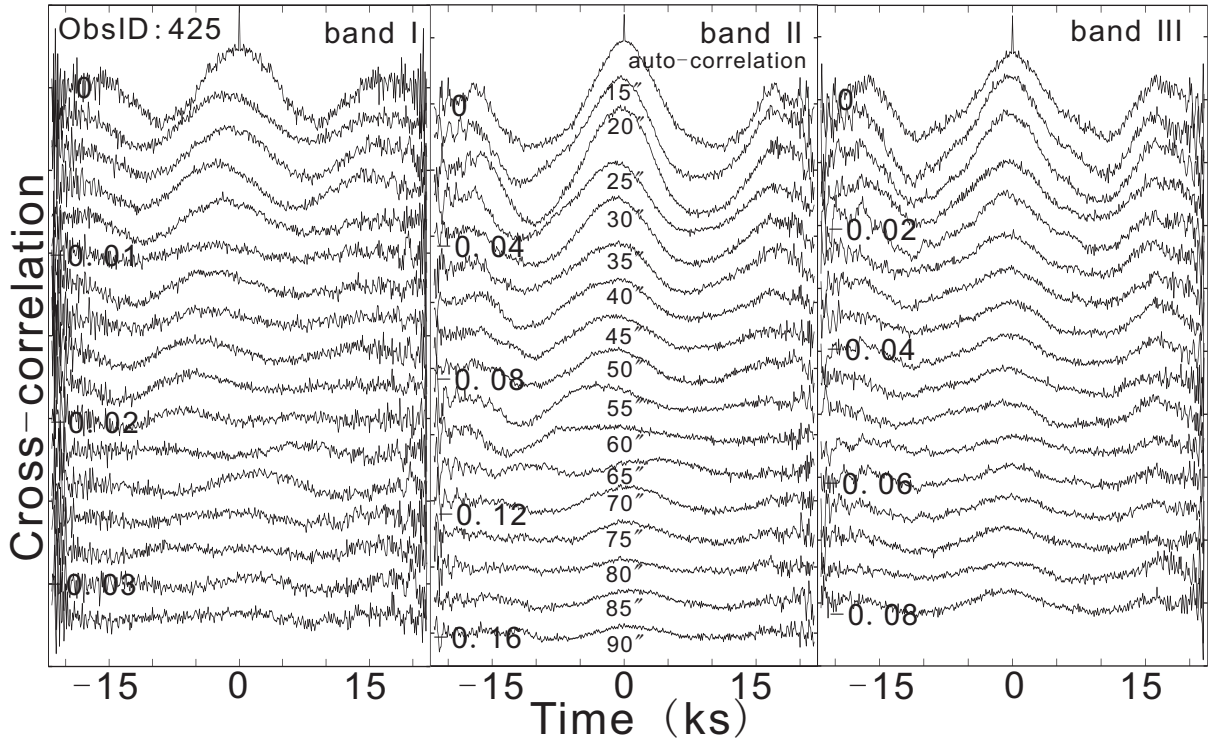}\\
\caption{Cross-correlation curves of ObsID425. The top curve in each
panel is the autocorrelation of the light curve of the source; all
other curves are the cross-correlation curves from 15$^{\prime
\prime}$ to 90$^{\prime \prime}$ with a step of 5$^{\prime \prime}$.
For clarity, the cross-correlation coefficients have been lowered by
a same amount successively for each curve.}\label{fig:ccf425}
\end{center}
\end{figure}

\begin{figure}
\begin{center}
\includegraphics[angle=0,scale=0.7]{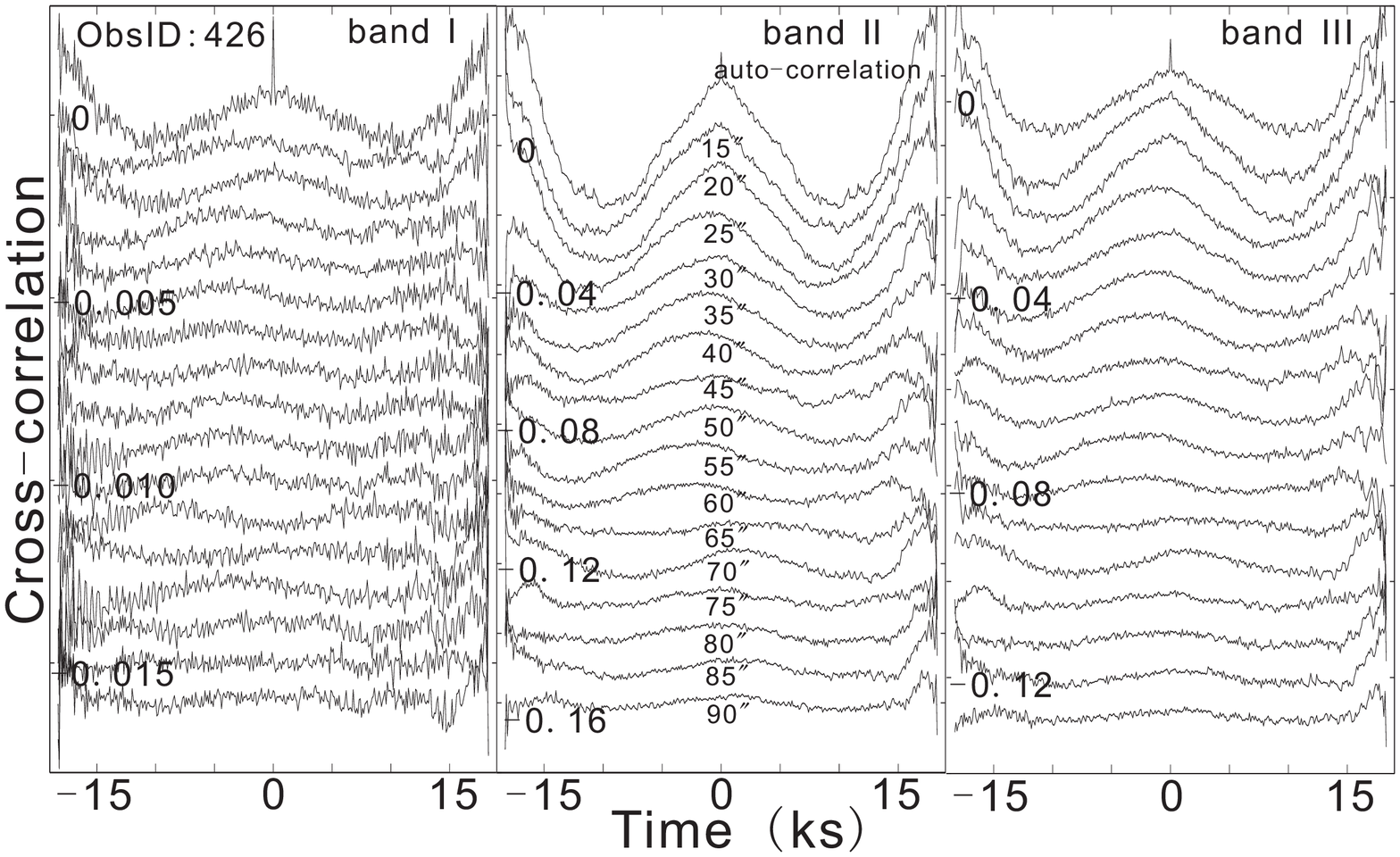}\\
\caption{Cross-correlation curves of ObsID426.}\label{fig:ccf426}
\end{center}
\end{figure}

\begin{figure}
\begin{center}
\includegraphics[angle=0,scale=0.7]{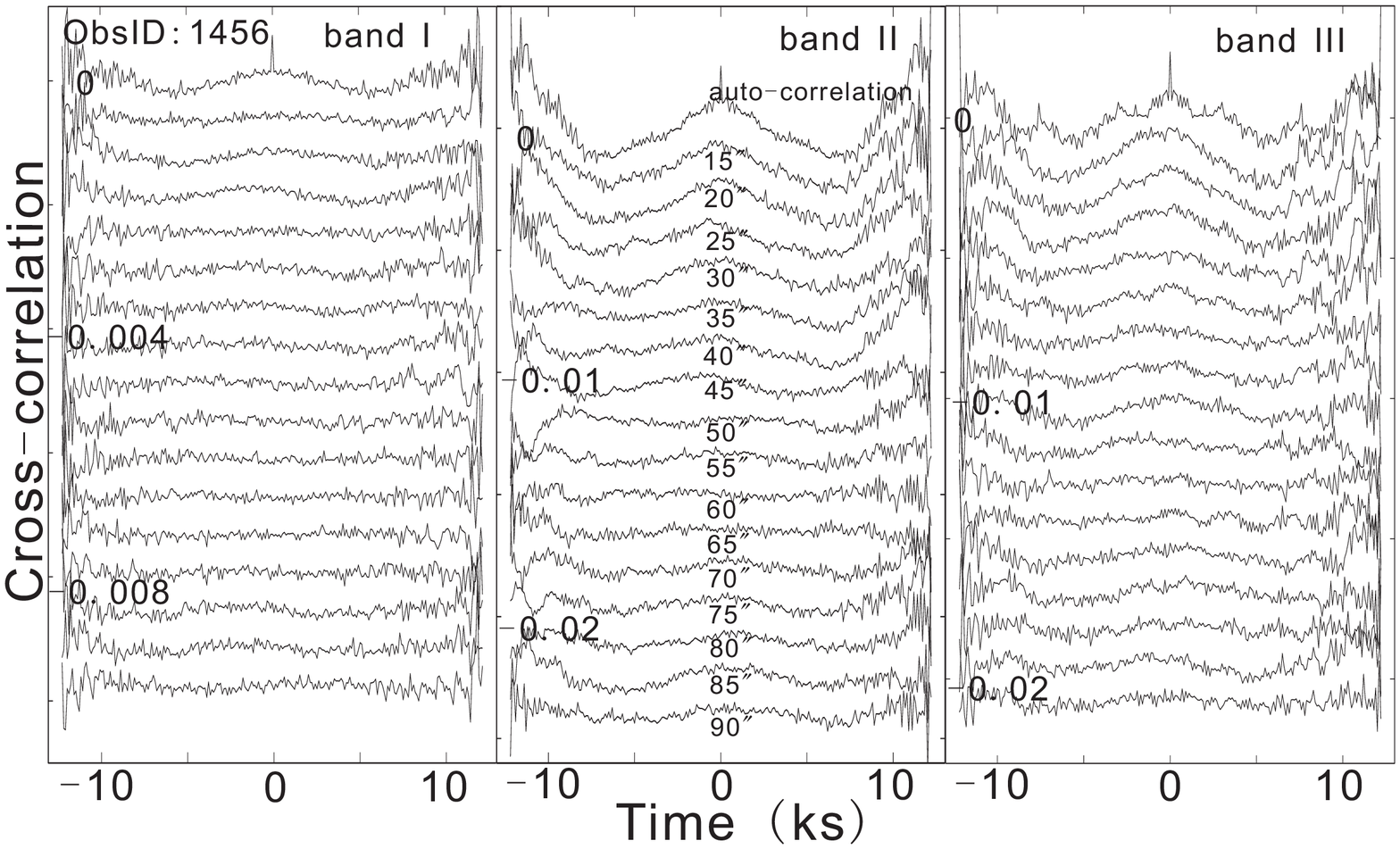}\\
\caption{Cross-correlation curves of ObsID1456.}\label{fig:ccf1456}
\end{center}
\end{figure}

\begin{figure}
\begin{center}
  \includegraphics[angle=0,scale=.8]{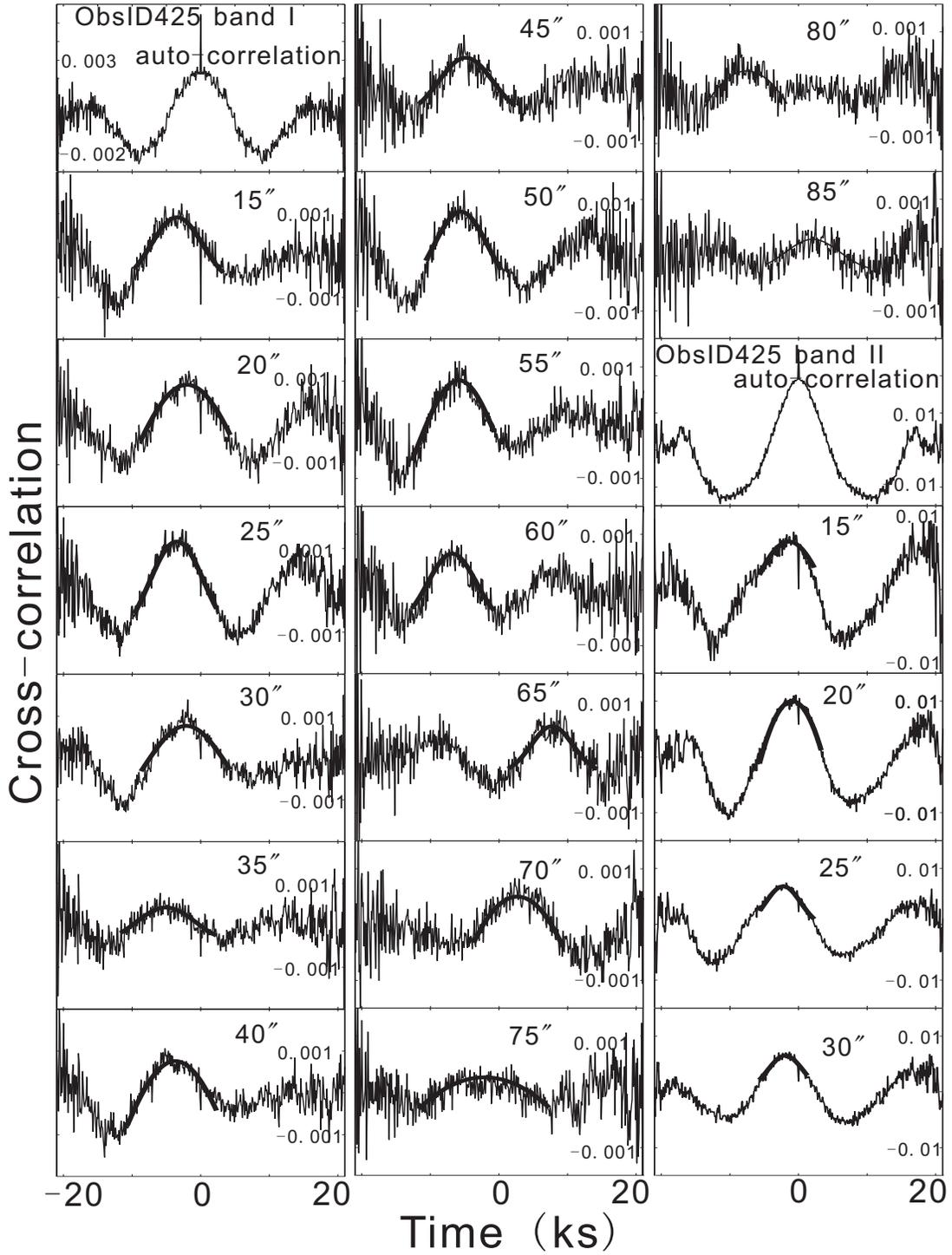}\\
  \caption{Cleaned cross-correlation curves used in this work.
The fitting results of the peaks are shown as the solid lines, with
a simple Gaussian function.}\label{fig:ccf1}
\end{center}
\end{figure}

\begin{figure}
\begin{center}
  \includegraphics[angle=0,scale=.8]{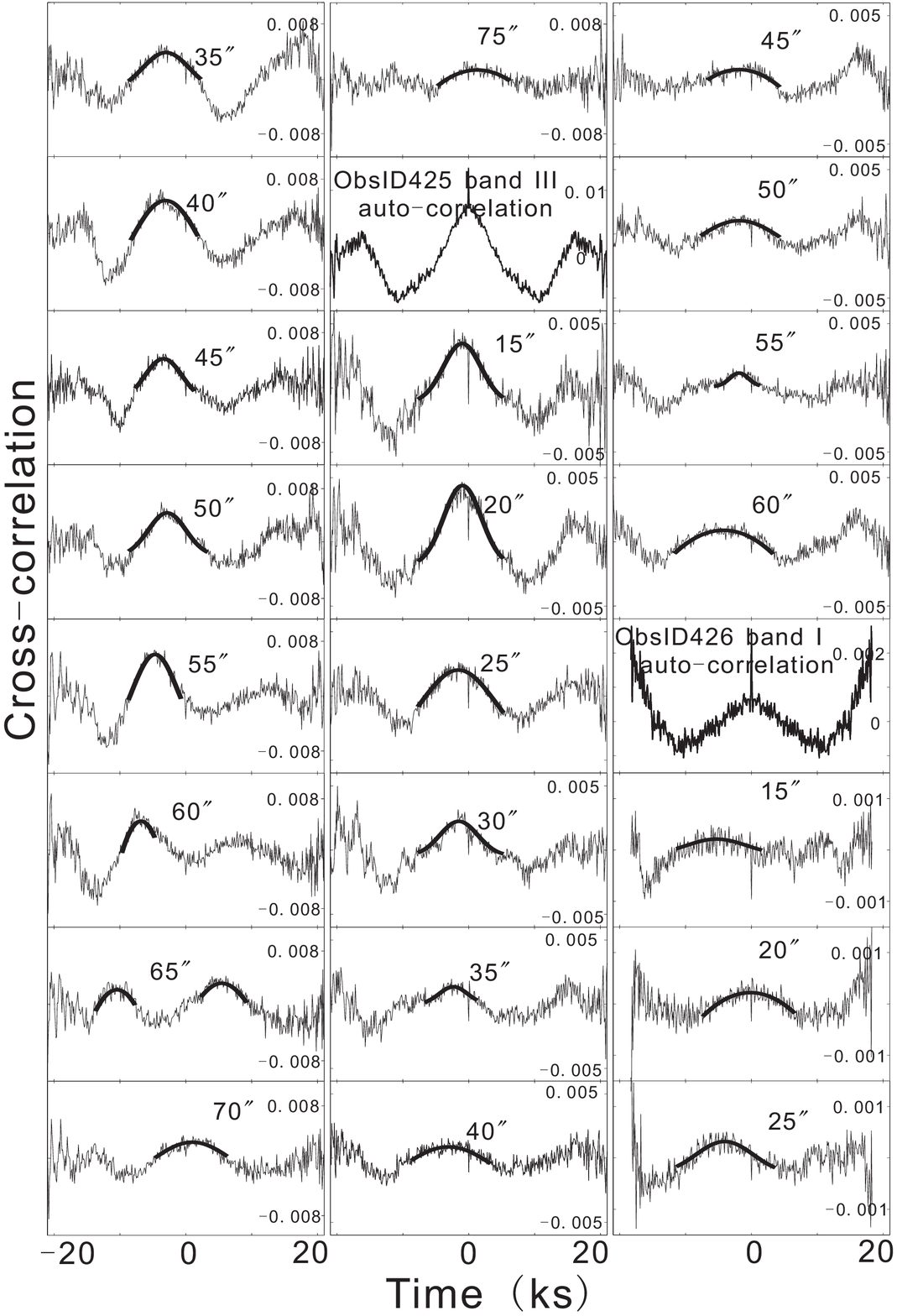}\\
  \caption{Fig.5 continued.}\label{fig:ccf2}
\end{center}
\end{figure}

\begin{figure}
\begin{center}
  \includegraphics[angle=0,scale=.8]{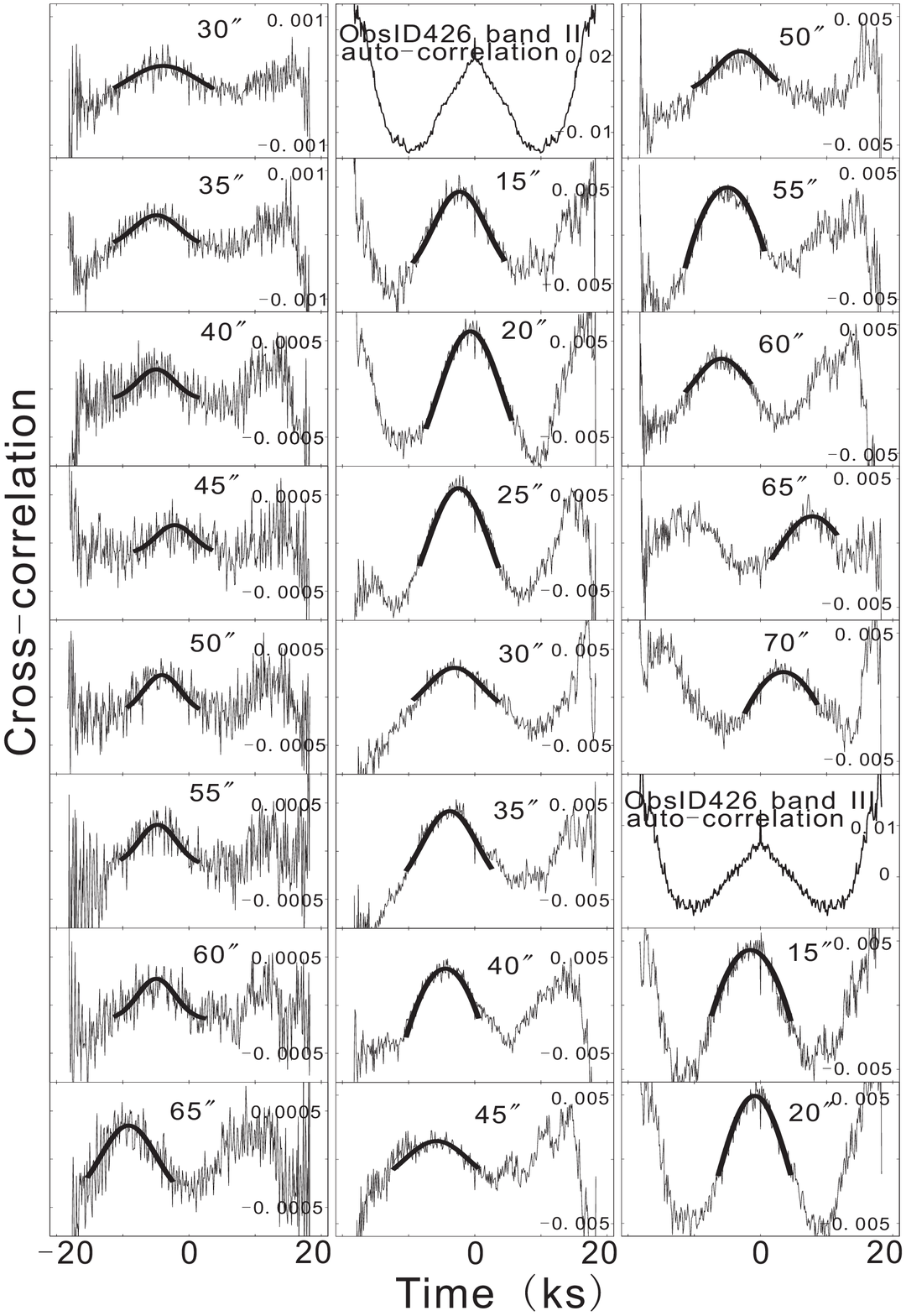}\\
  \caption{Fig.5 continued.}\label{fig:ccf3}
\end{center}
\end{figure}

\begin{figure}
\begin{center}
  \includegraphics[angle=0,scale=.8]{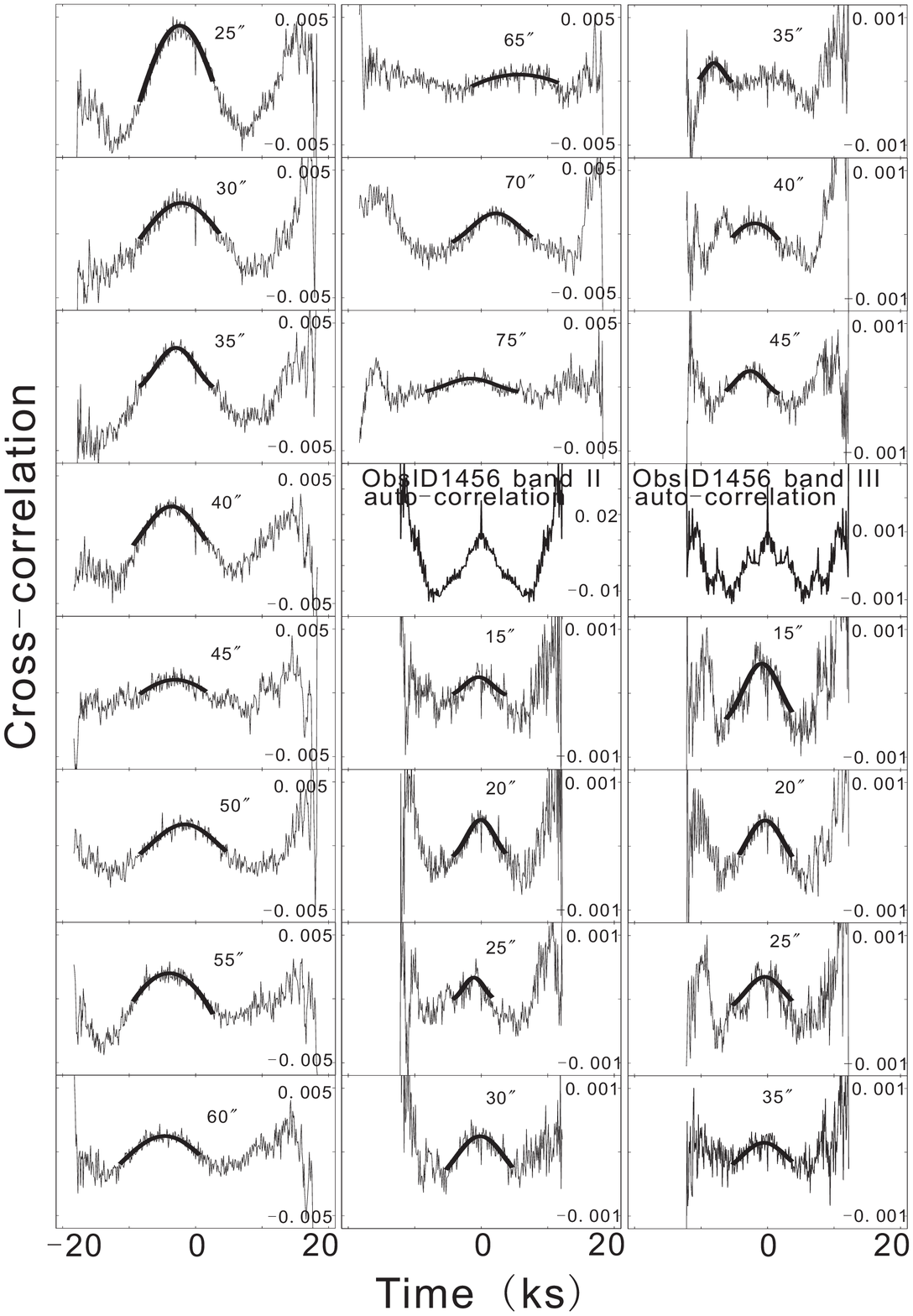}\\
  \caption{Fig.5 continued.}\label{fig:ccf4}
\end{center}
\end{figure}

\begin{figure}
\begin{center}
  \includegraphics[angle=0,scale=.9]{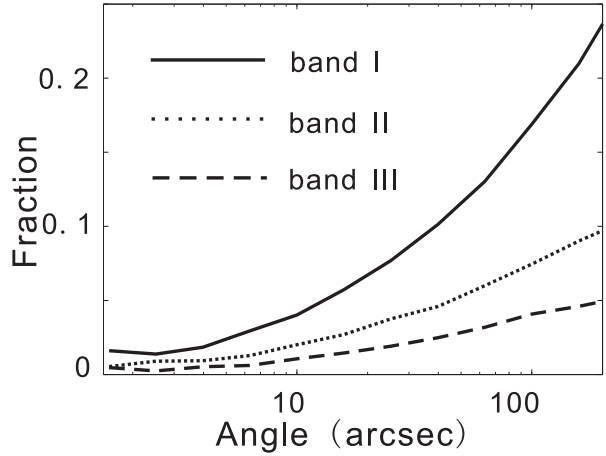}\\
  \caption{Simulated
fractions of multiple-scattered photons to the total halo photons
within each energy band as functions of the observational angle. The
parameters we used are $\tau_\mathrm{sca} = 2$ at \emph{E} = 1 keV
and \emph{a} = 0.1 $\mu$m. The solid line shows the fraction of
multiple-scattered photons for band I (below 3 keV). The dotted line
shows the fraction of multiple-scattered photons for band II (3 to 5
keV). The dashed line shows the fraction of multiple-scattered
photons for band III (above 5 keV).}\label{fig:multi}
\end{center}
\end{figure}

\begin{figure}
\begin{center}
  \includegraphics[angle=0,scale=.9]{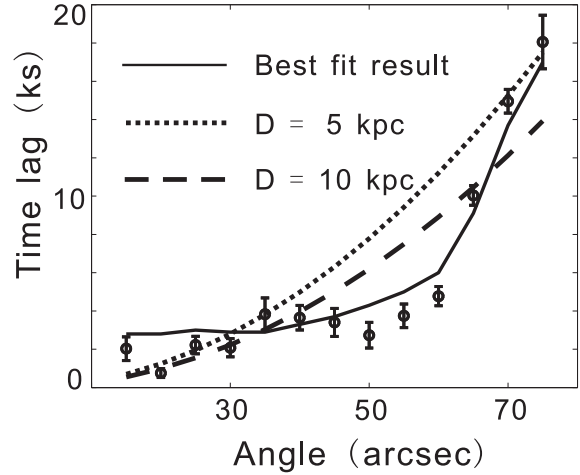}\\
  \caption{Observed time lag vs. observational angles.
  The solid line shows the best-fit result obtained in Section 4.3.
  The dashed line shows a model assuming a
dust wall located at 1.7 kpc from the Sun and a distance of 5 kpc to
the source. The dotted line shows the same model with a distance of
10 kpc to the source.}\label{fig:fit}
\end{center}
\end{figure}

\begin{figure}
\begin{center}
  \includegraphics[angle=0,scale=.9]{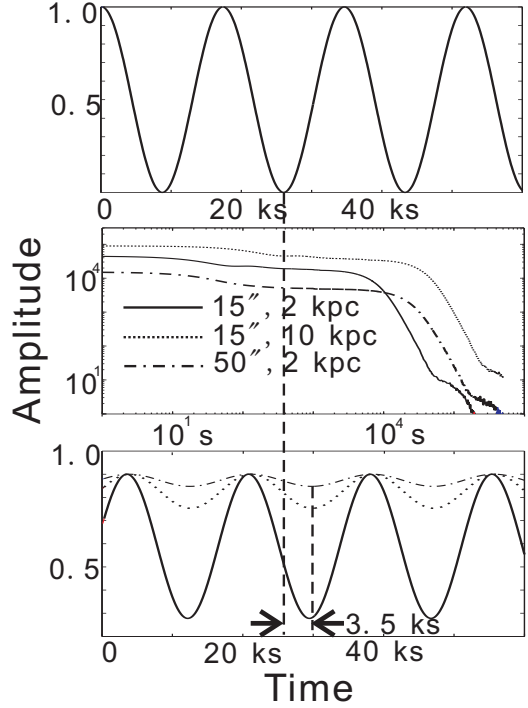}\\
  \caption{Convolution process with different parameters. The top panel
shows a sinusoidal wave, representing the light curve of the source.
The middle panel shows three curves of $\mathit{R(\phi, t)}$ in
logarithm scale with a uniform dust distribution. The solid line
shows the $\mathit{R(\phi, t)}$ with the parameter of $\mathit{D} =
2$ kpc and $\phi = $15$^{\prime \prime}$, the dotted line shows the
$\mathit{R(\phi, t)}$ of $D = 10$ kpc and $\phi$ = 15$^{\prime
\prime}$, and the dashed line shows the $\mathit{R(\phi, t)}$ of $D
= 2$ kpc and $\phi = 50$$^{\prime \prime}$. The bottom panel shows
the result of convolution with these three different
$\mathit{R(\phi, t)}$. These three curves are treated as the light
curves of halo at different observational angles. The vertical axis
of the bottom panel is normalized to unity for clarity. A lag of 3.5
ks is seen in these light curves directly.}\label{fig:convection}
\end{center}
\end{figure}


\begin{figure}
\begin{center}
  \includegraphics[angle=0,scale=.9]{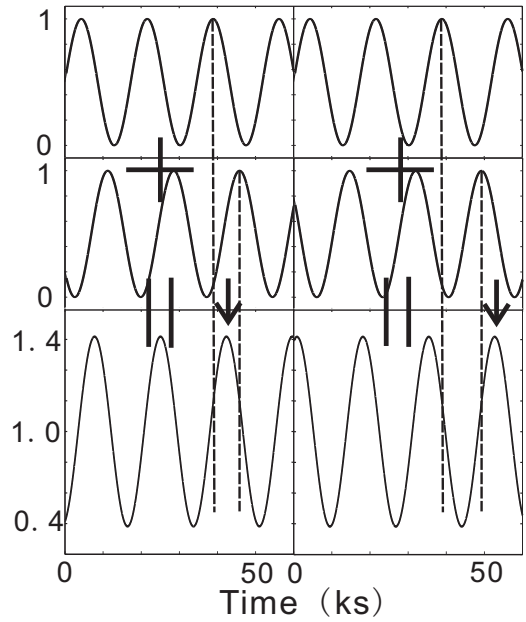}\\
  \caption{Illustration of the sum of two sinusoidal functions. The top
panel represents the light curve of $L(t) \otimes R_{u}(t)$, the
middle panel represents the light curve of $L(t) \otimes R_{w}(t)$.
The left part of the middle panel has a lag of less than $T/2$ with
respect to the top panel, and the right part has a lag of larger
than $T/2$ with respect to the top panel. The bottom panel shows the
sum of the above two panels. The arrows show the lag of the summed
curves.}\label{fig:sum}
\end{center}
\end{figure}

\begin{figure}
\begin{center}
  \includegraphics[angle=0,scale=.9]{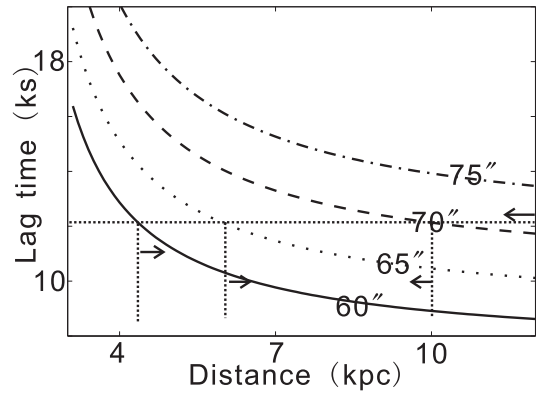}\\
  \caption{Geometrical lag time caused by the dust wall distribution as a
  function of the distance
  of the point source. The distance of dust wall is assumed to be 1.7 kpc for all cases.
  The solid line
  shows the lag time at 60$^{\prime \prime}$, the dotted line shows the lag time at
  65$^{\prime \prime}$, the dashed line
  shows the lag time at 70$^{\prime \prime}$, and the dashed dotted line shows the
  lag time
  at 75$^{\prime \prime}$. The four arrows indicate the distance upper
  or lower limits from those four angles.
  Finally, a distance range of [6, 10] kpc is obtained for Cyg X-3.
  }\label{fig:distance}
\end{center}
\end{figure}

\begin{figure}
\begin{center}
  \includegraphics[angle=0,scale=.9]{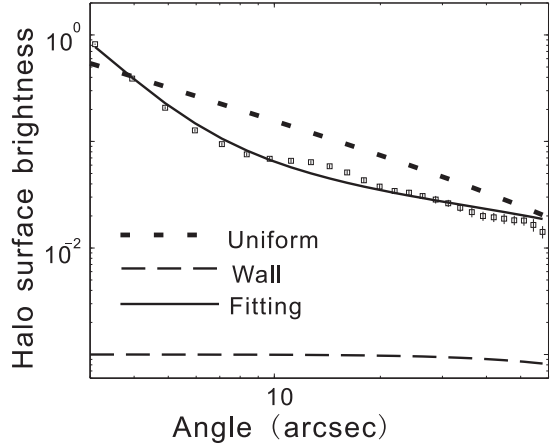}\\
  \caption{Halo surface brightness of Cyg X-3. The dotted line
  shows the predicted surface brightness of a uniform
distribution, and the dashed line shows the predicted surface
brightness caused by Cyg OB2. The total $\mathit{N_{H}}$ used here
is $3.0 \times 10^{22}$ $\mathrm{cm^{-2}}$, and the fraction of the
dust existing in Cyg OB2 is 7$\%$, according to the result obtained
with the cross-correlation method. The solid line shows the fitting
result of a three-components dust distribution. The column density
$\mathit{N_{H}}$ of the uniform distribution, the dust wall and the
dust near the source (at \emph{x} $>$ 0.99) are 7.0 $\pm$ 0.3
$\times 10^{21}$ $\mathrm{cm^{-2}}$, 32.0 $\pm$ 1.7 $\times 10^{21}$
$\mathrm{cm^{-2}}$ and 3.04 $\pm$ 0.06 $\times 10^{21}$
$\mathrm{cm^{-2}}$, respectively. The total $\mathit{N_{H}}$ from
the fitting is consistent with the result of Predehl and Schmitt
(1995). }\label{fig:surface}
  \end{center}
\end{figure}

\begin{figure}
\begin{center}
  \includegraphics[angle=0,scale=0.8]{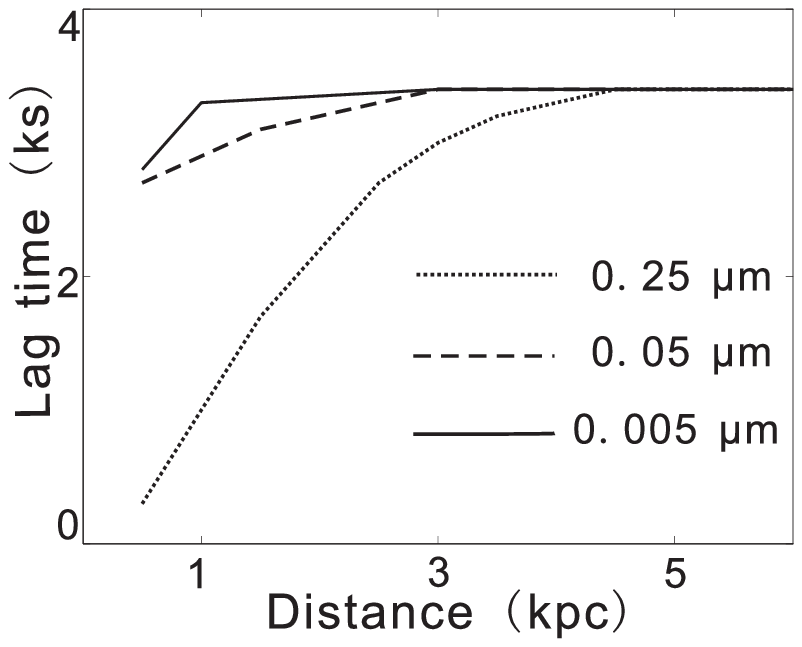}\\
  \caption{Lag time of $H_{u}(t)$ of 15$^{\prime \prime}$ vs. the distance of the source.
  The solid line shows the lag time of $H_{u}(t)$ with a dust
  radius of 0.005 $\mu$m, the dashed line shows the lag time of $H_{u}(t)$ with a dust
  radius of 0.05 $\mu$m, and the dotted line shows the lag time of $H_{u}(t)$ with a dust
  radius of 0.25 $\mu$m. The lag time approaches to 3.5 ks when the
  distance of the source exceeds 5 kpc for any dust radius.}\label{fig:dustmodel}
\end{center}
\end{figure}


\begin{thebibliography}{}
 \bibitem{Avni} Avni, Y., 1976, ApJ, 210, 642
 \bibitem{Di}  Dickey, J.M., 1983, ApJ, 273, L71
 \bibitem{Hu} Hu, J., Zhang, S. N., \& Li, T. P. 2004, ApJ, 614, L45
 \bibitem{Hump} Humphreys, R. M. 1978, ApJS, 38, 309
 \bibitem{Hutchings} Hutchings, J. B. 1981, PASP, 93, 50
 \bibitem{Kn} Kn$\mathrm{\ddot{o}}$dlseder, J. 2000, A\&A, 360, 539
 \bibitem{Kn} Kn$\mathrm{\ddot{o}}$dlseder, J. 2003, in IAU Symp. 212,A Massive Star Odyssey fromMain Sequence to
Supernova, ed. K. A. van der Hucht, A. Herrero, \& C. Esteban (San
Francisco: ASP), 505
 \bibitem{Ling} Ling, Z. X., Zhang, S. N., Xiang, J. G., \& Tang, S. C. 2009,
 ApJ, 690, 224
 \bibitem{Matri} Matri, J., Paredes, J. M., \& Peracaula, M. 2001,
 A\&A, 375, 476
 \bibitem{Massey} Massey, P., \& Thompson, A. B. 1991, AJ, 101, 1408
\bibitem{MathisLee} Mathis, J. S., \& Lee, C. -W. 1991, ApJ, 376,
490
 \bibitem{Mathis} Mathis, J. S., Rumpl, W., \& Nordsieck, K. H. 1977, ApJ, 217,
 425 (MRN)
\bibitem{Milgrom} Milgrom, M., 1979, A\&A, 76, L3
 \bibitem{Overbeck} Overbeck, J. W. 1965, ApJ, 141, 864
 \bibitem{Predehl2} Predehl, P., Burwitz, V., Paerels, F., \& Tr$\mathrm{\ddot{u}}$mper, J. 2000, A\&A, 357, L25
 \bibitem{Predehl1} Predehl, P., \& Schmitt, J. H. M. M. 1995, A\&A, 293, 889
 \bibitem{Rolf} Rolf, D. P. 1983, Nature, 302, 46
 \bibitem{Smith1998} Smith, R. K., \& Dwek, E. 1998, ApJ, 503, 831
 \bibitem{Smith} Smith, R. K., Edgar, R. J., \& Shafer, R. A. 2002, ApJ, 581, 562
 \bibitem{Thompson} Thompson, T. W., \& Rothschild, R. E. 2009, ApJ,
 691, 1744
 \bibitem{Torres} Torres$-$DodGen, A. V., Carroll, M., \& Tapia, M. 1991, MNRAS, 249, 1

 \bibitem{Trumper} Tr$\mathrm{\ddot{u}}$mper, J., \& Sch$\mathrm{\ddot{o}}$nfelder, V. 1973, A\&A, 25, 445
 \bibitem{van de Hulst} van de Hulst, H.C. 1957, Light Scattering by Small
 Particle (New York: Dover)
 \bibitem{Vaughan04} Vaughan, S., et al. 2004, ApJ, 603, L5
 \bibitem{Vaughan06} Vaughan, S., et al. 2006, ApJ, 639, 323
 \bibitem{WD01} Weingartner, J. C., \& Draine, B. T. 2001, ApJ, 548, 296 (WD01)
 \bibitem{Xiang, Lee } Xiang, J. G., Lee, J. C., \& Nowak, M. A. 2007, ApJ, 660, 1309
 \bibitem{Xiang} Xiang, J. G., Zhang, S. N., \& Yao, Y. S. 2005, ApJ, 628, 769

 \bibitem{Yao} Yao, Y. S., Zhang, S. N., Zhang, X. \& Feng, Y. 2003, ApJ, 59, L43

\end{thebibliography}
\end{document}